\documentstyle[cite,cas,epsfig]{article}
\newcommand{\be}{\begin{equation}}
\newcommand{\ee}{\end{equation}}
\newcommand{\ba}{\begin{array}{c}}
\newcommand{\ea}{\end{array}}
\newcommand{\beqn}{\begin{eqnarray}}
\newcommand{\eeqn}{\end{eqnarray}}

\newcommand{\bi}{\begin{itemize}}
\newcommand{\ei}{\end{itemize}}

\newcommand{\cL}{{\cal L}}

\newcommand{\cO}{{\cal O}}

\newcommand{\rms}{\rm\scriptsize}
\def\tcf{$\tau$cF}

\begin{document}

%
%

\begin{titlepage}
\vspace{0.3in}
\begin{flushright}
{CERN-TH.7065/93}
\end{flushright}
\vspace*{2.7cm}
\begin{center}
{\Large \bf TAU PHYSICS PROSPECTS\\ AT THE TAU-CHARM FACTORY \\
  AND AT OTHER MACHINES \\ }

\vspace*{0.4cm}
A. Pich$^{*\dagger}$

Theory Division, CERN, CH-1211 Geneva 23

\end{center}
\vspace*{1.5cm}

\centerline{ABSTRACT}

\noindent The prospects for tau physics at future high-luminosity facilities
are briefly discussed. Although
important (and often complementary) contributions will be made from
other machines, the unique experimental environment near
threshold
makes the Tau-Charm Factory the best
experimental tool for $\tau$ physics.

\vspace*{2.2cm}
\begin{center}
{Talk given at the\\
Marbella Workshop on the Tau-Charm Factory \\
Marbella, Spain, June 1993}
\end{center}

\vspace*{2.5cm}

{\footnotesize\baselineskip 10pt \noindent
$^*$ On leave of absence from
Departament de F\'{\i}sica Te\`orica, Universitat de Val\`encia,
and IFIC, Centre Mixte Universitat de Val\`encia--CSIC,
E-46100 Burjassot, Val\`encia, Spain.}

{\footnotesize\baselineskip 10pt \noindent
$^\dagger$ Work supported in part
by CICYT (Spain), under grant No. AEN93-0234.}

\vfill
\begin{flushleft}
{CERN-TH.7065/93 \\
November 1993}
\end{flushleft}
\end{titlepage}

%

\title{TAU PHYSICS PROSPECTS AT THE TAU-CHARM FACTORY
AND AT OTHER MACHINES}
\author{A. Pich\thanks{On leave of absence from
Departament de F\'{\i}sica Te\`orica, Universitat de Val\`encia,
and IFIC, Centre Mixte Universitat de Val\`encia--CSIC,
E-46100 Burjassot, Val\`encia, Spain.}\ \thanks{Work supported in part
by CICYT (Spain), under grant No. AEN93-0234.}}
\institute{Theory Division, CERN, CH-1211 Geneva 23}
\maketitle \thispagestyle{plain}
\begin{abstract}
The prospects for tau physics at future high-luminosity facilities
are briefly discussed. Although
important (and often complementary) contributions will be made from
other machines, the unique experimental environment near
threshold
makes the Tau-Charm Factory the best
experimental tool for $\tau$ physics.
\end{abstract}
\pagestyle{plain}

The physics programme of the Tau-Charm Factory
(\tcf) was 
established in the
1989 SLAC workshop \cite{tcws}. At that time, the experimental knowledge
on the $\tau$ was quite poor \cite{BS88}, and the main part of the
physics community was
not giving much attention to that (still exotic) lepton.
The situation has completely changed since then.
ARGUS and CLEO have substantially increased their $\tau^+\tau^-$
data sample, and the four LEP collaborations have
demonstrated the potential of this machine for
making clean $\tau$ physics.
Moreover, a small, but very important, contribution has been
accomplished at the Beijing collider.
The growing interest on the $\tau$ particle has been reflected
in several specialized reviews
\cite{TAUREV,reviews},
which have emphasized the unique properties of this heavy lepton
for testing the Standard Model,
and in the first two workshops  \cite{orsay,ohio}
devoted entirely to the $\tau$.
The qualitative change of $\tau$ physics can be appreciated in
Table \ref{tab:improvements},
which compares the status of several $\tau$ measurements in the
1990 compilation of the Particle Data Group \cite{PDG90}
with the more recent world averages \cite{ohio,fernandez};
the main experimental sources of the improvements are also indicated.

Obviously,
our knowledge on the $\tau$ lepton properties is going to be
further improved in the next few years.
To get a proper feeling on the
quantitative impact
of the \tcf\ in this field,
one should analyze the precisions that can be reached with
present facilities.
Moreover, the possible contributions of other future machines
should also be considered.
Figure \ref{fig:colliders} shows the energy and luminosity of present and
future $e^+e^-$ colliders
in the energy range 1 GeV $\leq E_{\mbox{\rms c.m.}}\leq$ 100 GeV.
All machines above $\sqrt{s} = 2 m_\tau$ produce $\tau$'s and can
make significant contributions to $\tau$ physics.

\begin{table}[h]
\caption{Recent improvements in $\tau$ physics.
[(a) 95\% C.L.;   (b) 90\% C.L.]}
\label{tab:improvements}
\begin{center}
\begin{tabular}{|c|c|c|c|}
\hline
Parameter & 1990 \protect\cite{PDG90} &
  1993 \protect\cite{ohio,fernandez} & Experiment
\\ \hline
$m_\tau$ \, (MeV) & $1784.1^{+2.7}_{-3.6}$ & $1777.0\pm 0.3$ & BES
\\
$m_{\nu_\tau}$ \, (MeV) & $< 35$ \quad (a) & $< 31$ \quad (a)
&  ARGUS + BES
\\
$\tau_\tau$ \, (fs) & $303\pm 8$ & $294.7\pm 3.0$ & LEP, CLEO
\\
$B_e$ \, (\%) & $17.9\pm 0.4$ & $17.88\pm 0.15$ & LEP, CLEO, ARGUS
\\
$B_\mu$ \, (\%) & $17.8\pm 0.4$ & $17.42\pm 0.17$ & LEP, ARGUS
\\
$B(h^-\nu_\tau)$ \, (\%) & $12.6\pm 1.2$ & $12.04\pm 0.24$ & LEP, ARGUS
\\
$B_1$ \, (\%) & $86.13\pm0.33$ & $85.23\pm 0.19$ ? & LEP ($84.80\pm 0.23$)
\\
$B(3\pi^\pm 2\pi^0\nu_\tau)$ \, (\%) & -- & $0.54\pm 0.10$ & CLEO
\\
$B(\pi^-\pi^0\eta\nu_\tau)$ \, (\%) & $< 1.1$ \quad (a) &
   $0.17\pm 0.02$ & CLEO
\\
$B(\mu^-\gamma)$ & $< 5.5\times 10^{-4}$ \quad (b) &
  $< 4.2\times 10^{-6}$ \quad (b) & CLEO
\\
$B(e^-e^+e^-)$ & $< 3.8\times 10^{-5}$ \quad (b) &
  $< 3.5\times 10^{-6}$ \quad (b) & CLEO
\\
$\rho_e$ & $0.64\pm 0.06$ & $0.717\pm 0.038$ & ARGUS
\\
$\rho_\mu$ & $0.84\pm 0.11$ & $0.762\pm 0.046$ & ARGUS
\\
$\xi_l$ & -- & $0.90\pm 0.18$ & ARGUS
\\
$h_{\nu_\tau}$ & -- & $-1.25\pm0.26$ & ARGUS
\\
$\tilde{d}_\tau^Z(M_Z)$ \, (e cm) & -- & $< 3.7\times 10^{-17}$ & LEP
\\
$\mu(\nu_\tau)$ \, ($\mu_B$) & $< 4\times 10^{-6}$ \quad (b) &
  $< 5.4\times 10^{-7}$ \quad (b) & BEBC
\\
$B(Z\to\tau\tau)$\, (\%) & $3.33\pm0.13$ & $3.36\pm0.02$ & LEP
\\
$B(Z\to\tau e)$ & -- & $< 1.3 \times 10^{-5}$ \quad (a) & LEP
\\
$B(Z\to\tau\mu)$ & -- & $< 1.9 \times 10^{-5}$ \quad (a) & LEP
\\ \hline
$\alpha_s(m_\tau)$ & 0.12 -- 0.41 & $0.35\pm0.03$ & LEP, CLEO, ARGUS
\\ \hline
$\sum_{\mbox{\rms excl.}} B_i$ \, (\%) & $93.5\pm2.4$ &
$\ba 100.3\pm1.3 \\ 91.0\pm3.3 \ea$ ? &
$\ba \mbox{\rm ALEPH} \\ \mbox{\rm ARGUS} \ea$
\\ \hline
\end{tabular}
%
\end{center}
\end{table}

\begin{figure}[htb]
\caption{Luminosity/Energy plot of present and
future $e^+e^-$ colliders,
in the energy range 1 GeV
$\leq E_{\mbox{\protect\rms c.m.}}\leq$
100 GeV.}
\label{fig:colliders}
\end{figure}

The different running energies have their own advantages and problems;
thus, a given collider can be very good for measuring some
parameter and totaly insensitive to other properties of the $\tau$.
The luminosity is clearly an important ingredient, but not always
the decisive one.
There are three energy regions worth while to be considered:
the $\tau^+\tau^-$ threshold (BEPC, \tcf),
the $\Upsilon$ region (DORIS, CESR, BF) and the $Z$ peak (LEP, ZF).

LEP has the great advantage of producing $\tau$'s with a sizeable
boost and  low backgrounds. It is obviously the best
machine for lifetime measurements. Adding the 1990--1992 data sample
of the four LEP experiments ($2\times 10^5$ $\tau$ pairs) an
accuracy of about 3 fs has been obtained \cite{kounine}:
$\tau_\tau|_{\mbox{\rms LEP}} = 293.5\pm2.8$ fs.
In spite of having accumulated a much larger statistics
($2\times 10^6$ $\tau$ pairs), the CLEO result \cite{fernandez},
$\tau_\tau|_{\mbox{\rms CLEO}} = 296\pm10$ fs,
is less precise;
a smaller average flight path and larger hadronic backgrounds
make difficult to achieve a better sensitivity \cite{weinstein}.
At threshold, to measure the lifetime is clearly not possible.

The $\tau$-mass measurement requires completely different experimental
conditions.
With only 7 ($\tau^+\tau^-\to e^\pm\mu^\mp + 4\nu$) events,
taken at threshold, BES \cite{BES} has been able to achieve an unbelievable
accuracy of 0.5 MeV: $m_\tau = 1776.9\pm 0.5$ MeV.
A recent update, including more channels, has further improved
the precision to 0.3 MeV \cite{fernandez}.
ARGUS and CLEO, with many orders of magnitude more data,
have only
reached precisions of 2.8 and 1.8 MeV, respectively \cite{fernandez},
using clever kinematic tricks.

Between these two extreme cases (lifetime and mass), nearly
all  aspects of $\tau$ physics
(except the $Z\tau^+\tau^-$ couplings)
can be addressed in any of the three
energy regions;
with slightly different strategies, because of the different
kinematics (and background) conditions.
Nevertheless, the sensitivity to a given measurement can change with
the centre-of-mass energy
(or with the symmetric/asymmetric configuration of the machine
\cite{weinstein}).
The potential information provided by different facilities should
then be regarded as complementary.
Tau physics can and should be done at any (existing or planned)
$e^+e^-$ collider
above threshold.

Present experiments can still increase their sensitivities
and better (in some cases new) results should be expected
in the near future; however, they are soon going to reach
their systematic limits \cite{kounine,weinstein,cowen,gomez,diberder}.
Further improvements in $\tau$ physics require new high-precision machines
working in the high-statistics regime.
As shown in Fig. \ref{fig:colliders} three high-luminosity
colliders have been proposed, one in each of the three
energy regions discussed before: the \tcf, the $B$ Factory (BF)
and the $Z$ Factory (ZF).
Table \ref{tab:data_samples} shows the number of $\tau^+\tau^-$
pairs that would be produced by these facilities.
The highest production rate corresponds to the \tcf,
but the differences are not very large. 
Note that these numbers would further increase
if higher luminosities are achieved.
A next generation BF could reach
$\cL = 10^{34} \mbox{cm}^{-2} \mbox{s}^{-1}$
and a high-luminosity \tcf\ with
$\cL = 5\times 10^{33} \mbox{cm}^{-2} \mbox{s}^{-1}$
has already been considered at this workshop \cite{jasper,duff}.

\begin{table}[hbt]
\caption{Comparison of $\tau$ yearly data samples  at the
Z, B and $\tau$-charm factories.}
\label{tab:data_samples}
\begin{center}
\begin{tabular}{|c||c|c|c|}
\hline
Collider & ZF & BF & \tcf
\\ \hline
Luminosity ($\mbox{cm}^{-2} \mbox{s}^{-1}$) &
$2\times 10^{32}$ & $10^{33}$ & $10^{33}$
\\ \hline
$N_{\tau^+\tau^-}$ &  $0.3\times 10^7$ & $0.9\times 10^7$ &
 $\ba   0.5\times 10^7 \; \mbox{\rm (3.57 GeV)}
\\
 2.4\times 10^7 \;  \mbox{\rm (3.67 GeV)}
\\
 3.5\times 10^7 \; \mbox{\rm (4.25 GeV)}  \ea $
\\ \hline
\end{tabular}
\end{center}
\end{table}

The larger statistics is, however, not the main advantage of the \tcf.
With the sharp increase in statistics foreseen in these facilities,
the attainable precision will  be limited by
backgrounds and systematic errors.
To achieve precise --$\cO (0.1\%)$-- measurements
of $\tau$ branching ratios, for instance,
the normalization, detection efficiencies and backgrounds must each
be known at the 0.1\% level.
The decisive advantage
that makes the \tcf\ the best experimental tool for $\tau$
physics,  is
its capability of tightly controlling the backgrounds and systematic errors.

The threshold region provides a unique environment, where backgrounds
are both small and experimentally measurable.
By adjusting the beam energy above or below the $\tau^+\tau^-$ threshold,
the backgrounds can be directly measured, avoiding any need
for Monte Carlo simulations with their inevitable uncertainties.
Moreover, the data samples are very pure because they are free of
heavier flavour backgrounds.
$b$ contaminations are completely absent and,
running below the charm threshold,
one can collect $\tau$ events that are
completely free of $c$ backgrounds.

Furthermore, at the \tcf\ it would be possible --for the first time
at any machine-- to single-tag $\tau^+\tau^-$ events;
i.e. by observing the decay
of one  $\tau$, its partner would be cleanly tagged,
without any pre-selection of its decay mode.
This completely avoids normalization uncertainties, which
turns out to be crucial to achieve very precise measurements
of branching ratios.
Single-tagging requires a clean signature from a single $\tau$ decay.
With the capability of the \tcf\ to produce $\tau$'s without
heavy flavours contamination, there are several
distinct signatures \cite{gomez}: $e+E_{\mbox{\rms miss}}$,
$\mu+E_{\mbox{\rms miss}}$, and (at 3.57 GeV)
monochromatic-$\pi + E_{\mbox{\rms miss}}$.
Detailed studies indicate backgrounds of between
$10^{-3}$ (at 3.57 GeV) and $10^{-4}$ (at 3.67 GeV)
for the $e+E_{\mbox{\rms miss}}$ trigger, which has a
$\tau^+\tau^-$ event detection efficiency of 24\%.
For comparison, the typical background contaminations
in present samples of double-tagged $\tau^+\tau^-$ events
at $B$ or $Z$ Factory energies are about 5\%.
The lowest backgrounds achieved in LEP data are about 1\%,
while retaining adequate detection efficiency.
This corresponds to a $q\bar q$ rejection of $\approx 4\times 10^{-4}$,
to be compared with $\approx 10^{-5}$ at the \tcf.

  The threshold region has also several
kinematic advantages that result from the low particle velocities,
such as monochromatic spectra for two-body
decays.
Due to the Coulomb interaction, the $\tau^+\tau^-$
production cross-section has a finite value of 0.20 nb
\cite{perrotet}  at threshold,
which makes feasible to collect a copious sample of
$\tau^+\tau^-$ events almost at rest.
Running just above the $\tau^+\tau^-$ threshold at 3.57 GeV,
the one-prong decays
$\tau^-\to l^-\bar\nu_l\nu_\tau,\pi^-\nu_\tau,K^-\nu_\tau$
are kinematically separated (Fig. \ref{fig:particle}a).
In contrast, at higher energies the distributions
completely overlap (Fig. \ref{fig:particle}b)
and the particle identification requirements
--especially between $\pi$ and $K$-- are more severe \cite{gomez2}.
Measurements of $\tau$ branching ratios can then be done
much more precisely at threshold.

\begin{figure}[htb]
\caption{Momentum spectra from one-prong $\tau$ decays at
centre-of-mass energies of a) 3.57 GeV (1.5 MeV above the $\tau^+\tau^-$
threshold) and b) 10 GeV.}
\label{fig:particle}
\end{figure}

Figure \ref{fig:pex}
clearly illustrates the power of this kinematic constraint
to search for exotic
two-body decays like $\tau^-\to l^- X$
($l=e,\mu$; $X$ = Majoron, familon, flavon, \ldots),
which,
at threshold,
lead to the distinctive signature of monochromatic leptons (Fig.
\ref{fig:pex}a)\cite{concha}.  In contrast, the sensitivity is much weaker at
higher energies since the lepton spectrum from the $l^- X$ decay is
broad,  spreading over the full spectrum of the
standard $l^-\bar{\nu}_l\nu_{\tau}$ decay (Fig. \ref{fig:pex}b).
For $\tau^-\to e^- X$ ($\tau^-\to \mu^- X$), the
expected branching-ratio sensitivity at the \tcf\  is better than $10^{-5}$
($10^{-3}$)
for one-year's data \cite{concha}, to
be compared with the present limits of 1-2\%.  The experimental sensitivity
could be improved a further order-of-magnitude if the monochromator
optics is successful.  Sensitivity to still-lower
branching ratios would require improvements in the $\pi e$ rejection of the
\tcf\ detector, e.g. with a fast RICH.

\begin{figure}[htb]
\caption{The combined electron spectra (solid histograms) from the standard
decay  $\tau^-\to e^-\bar\nu_e\nu_\tau$
mixed with a hypothetical 1\%
two-body decay $\tau\to eX$  at   centre-of-mass energies
of a) 3.57 GeV  and b) 10 GeV \protect\cite{concha}.
The particle $X$ is a
massless Goldstone boson.  The dashed histograms show conventional (V-A)
fits to the combined spectra.  Each plot contains 200k
$\tau^-\to e^-\bar\nu_e\nu_\tau$,
corresponding to two months' data at 3.57 GeV. }
\label{fig:pex}
\end{figure}

At threshold, the experimental data require small radiative
corrections since the $\tau$'s are almost at rest and
radiation can only bring them closer to rest.
At higher-energies, the important role of initial-state
QED bremsstrahlung results in large radiative corrections,
which unavoidably introduce additional contributions
into the final errors.
This effect is particularly important for
high-precision measurements of the Michel parameters in
leptonic $\tau$ decays.
In the recent ARGUS measurement \cite{ARGUSxi} of the $\xi$ parameter,
$|\xi| = 0.90\pm 0.13\pm 0.10$,
initial-state radiation is the second
main source of uncertainty with a $\pm0.06$
contribution to the final systematic error
(background subtraction gives $\pm0.08$).
Furthermore, the Lorentz boost smears
the final charged-lepton energy spectrum
and causes the parameters $\rho$ and $\eta$ to be strongly correlated.
Clearly, the Lorentz structure of $\tau$ decays
can be more accurately analyzed in the threshold region,
where the \tcf\ operates.

  Another important advantage of the \tcf\ energy region is the existence
of two precisely-known energy points, the $J/\psi$ and $\psi'$, which
provide a very high-rate ($\sim 1$~kHz) signal to calibrate and
monitor the detector performance. This
allows for a tight control of the systematic errors,
making possible, for instance,
to maintain the momentum-scale error of the \tcf\ detector
below 0.1\% \cite{jasper}.

In addition to excellent momentum measurements, there are advantages
\cite{jasper}
for photon detection and particle identification near threshold.
Since particles are produced isotropically, the
detection inefficiency caused by charged and neutral pileup is
minimized. This is especially important for decays involving
several neutral particles.
The very good photon resolution of the \tcf\ detector would
increase the sensitivity to the $\tau$-neutrino mass,
by allowing an optimal use of high-multiplicity $\tau$ decays containing
neutrals, such as $\tau^-\to\nu_\tau 2\pi^-\pi^+ 2\pi^0$
\cite{cowen,gomez}.
Moreover, the kinematic limit of
particles from $\tau$ decays at rest is 1.8 GeV,
making the identification of $\pi$, $K$ and $p$ easier
than at higher-energy machines.

The unique purity of the \tcf\ data
would offer the opportunity to perform a high-precision
global analysis of (inclusive and exclusive) semi-leptonic
$\tau$-decay modes,
including kinematical quantities more subtle than the simple
invariant-mass distribution of the hadronic final state
\cite{diberder}.
The cleanness of the (bias-free) data,
the excellent $\pi/K$ separation
and the very good efficiency for neutral modes,
would enourmously simplify the QCD analysis of current spectral functions,
removing the Monte Carlo based unfolding procedure which is needed,
at higher energies, to correct for detector effects \cite{diberder}.

Thus, the \tcf\ experiment would
benefit from a very
high statistics, low and measurable backgrounds, and reduced
systematic errors.
The coincidence of all these features near threshold
creates
an ideal facility for precision $\tau$ studies.
Table \ref{tab:precision} gives an illustrative list of
expected sensitivities
\cite{gomez,gomez2,concha,stahl,FETSCHER,pich}
at the \tcf\ for some
typical $\tau$ parameters.
In all cases the improvements with respect to the present
precisions are substantial.
Note, that a BF (or ZF) with a similar data sample
could obviously reach a comparable statistical precision
for many of these observables.
However, for most measurements,
only the \tcf\ is likely to achieve comparable systematic errors,
thereby reaching the overall precisions indicated in
Table \ref{tab:precision}.

\begin{table}[htb]
\caption{Examples of the
expected precision at the \tcf\ for some $\tau$ parameters.
 Some improved estimates from preliminary analyses are indicated within
  brackets.}
\label{tab:precision}
\begin{center}
\begin{tabular}{|c|c|c|}
\hline
Parameter & Present accuracy & \tcf\ sensitivity
\\ \hline
$m_\tau$ & 0.3 MeV & 0.1 MeV
\\
$m_{\nu_\tau}$   & $< 31$ MeV & 2 MeV
\\
$B_{e,\mu}$ & 1\% & 0.1\%
\\
$B_\pi$ & 3\% & 0.1\%
\\
$B_K$ & 34\% (12\%) & 0.5\%
\\
$|g_\tau/g_\mu|$ & 0.6\% & 0.1\%
\\
$|g_\mu/g_e|$ & 0.6\% & 0.1\%
\\
$\rho_{e,\mu}$ & 6\% & 0.3\% (0.03\%)
\\
$\xi_l$ & 20\% & 3\% (0.2\%)
\\
$h_{\nu_\tau}$ & 22\% & 0.3\%
\\
$\eta_\mu$, $\delta_l$ & -- & $\pm 0.03$ ($\pm 0.002$)
\\
$\xi'_\mu$ & -- & 15\%
\\
$B(\tau^-\to\pi^-\eta\nu_\tau)$ & $< 3\times 10^{-4}$ & $10^{-6}$
\\
$B(\tau^-\to l^- X)$ & $< 2\% $ & $10^{-5}$
\\
$B(\tau^-\to 3 l^\pm)$ & $< 10^{-5}$ & $10^{-7}$
\\
$B(\tau^-\to\mu^-\gamma)$ & $< 4\times 10^{-6}$ & $10^{-7}$
\\
$a_\tau^\gamma$ & $< 0.1$ & 0.001
\\
$d_\tau^\gamma$ & $< 6\times 10^{-6}$ e cm & $10^{-7}$ e cm
\\ \hline
\end{tabular}
\end{center}
\end{table}

\section*{Acknowledgements}
   Being a theorist, I feel quite embarrassed talking about
experimental acceptances, backgrounds, efficiencies, resolutions and so on.
I am indebted to my experimental colleagues in the $\tau$-physics
working group for their patience in helping me with this task.
I would like to thank J.J. G\'omez-Cadenas and J. Kirkby
for many discussions and useful comments on the manuscript.

\end{document}